\documentclass[twocolumn,showpacs,amsmath,amssymb,pra,intlimits]{revtex4}

\usepackage{graphicx}
\usepackage{dcolumn}
\usepackage{bm}
\usepackage{amssymb}
\usepackage{bbm}
\usepackage{calc}

\newcommand{\ii}{ {\rm i} }

\newcommand{\eq}[1]{Eq.~(\ref{#1})}

\begin{document}

\title{First-order strong field approximation for high-order harmonic generation}
\author{Ariel Gordon}
\email{gariel@mit.edu}
\author{Franz X. K\"artner}%

\affiliation{%
Department of Electrical Engineering and Computer Science and
Research Laboratory of Electronics, Massachusetts Institute of
Technology, Cambridge, Massachusetts 02139
}%

\begin{abstract}
Recently it was shown [A. Gordon and F. X. K\"artner, Phys. Rev.
Lett. \textbf{95}, 223901 (2005)] that the strong field
approximation (SFA) for high-order harmonic generation (HHG) is
significantly improved when the SFA wave function is used with the
acceleration rather than the length form of the dipole operator. In
this work it is shown that using the acceleration form upgrades the
SFA from zeroth-order to first-order accuracy in the binding
potential. The first-order correct three-step model
($\rm1^{st}$-order TSM) obtained thereby is systematically compared
to its standard zeroth-order counterpart ($\rm0^{th}$-order TSM) and
it is found that they differ significantly even for energetic
electrons. For molecules (in the single-electron approximation), the
$\rm0^{th}$-order and the $\rm1^{st}$-order TSMs in general disagree
about the connection between the orbital symmetry and the positions
of the minima in the HHG spectrum. At last, we briefly comment on
gauge and translation invariance issues of the SFA.
\end{abstract}


\maketitle

\section{Introduction}
The strong field approximation (SFA) \cite{Keldysh, Faisal, Reiss}
is a key technique in the study of interactions of matter with
intense laser fields. In particular, the SFA is used to describe
high-order harmonic generation (HHG) \cite{Becker89,Corkum,
Lewenstein,IvanovBrabec}. The three-step model (TSM) \cite{Corkum,
Lewenstein,IvanovBrabec}, which is based on the SFA, has proven very
successful describing much of the experimental behavior. The best
known examples are the cutoff formula \cite{Corkum,
Lewenstein,IvanovBrabec} and time-frequency structure of the HHG
signal \cite{Salieres}, which led to the prediction of attosecond
pulses \cite{atto}. Reviews can be found in
Refs.~\cite{BrabecReview, KnightReview}.

It has been long known however that the TSM is incorrect by 1-2
orders of magnitude predicting the spectral intensity of HHG in
atomic hydrogen, as found from comparisons with numerically-exact
results \cite{Tempea, BrabecReview}. For the $\rm  H_2^+$ ion the
TSM gives a spectrum which is \emph{many} orders of magnitude away
from numerically-exact results \cite{GKPRL}, and the shape of the
spectrum can be heavily distorted \cite{GKPRL, Bandrauk1}.

HHG experiments are now coming to the point where more accurate
theory is needed. The race towards harnessing HHG for a coherent
short wavelength source \cite{KrauszWaterWindow, Murnane1997,
KrauszNature05} can benefit from quantitative theoretical estimates
of the HHG efficiency, which to date are very scarce
\cite{BrabecEffi}. The recent orbital imaging experiment
\cite{CorkumNature} uses the HHG spectrum to infer the structure of
a molecular orbital, which also requires quantitatively reliable
theory, capable of giving a precise description of the shape of the
HHG spectrum.

In a recent theoretical work \cite{GKPRL} we proposed a modified
version of the TSM, where the SFA wavefunction is used with the
dipole operator in the acceleration rather than the length form.
Comparison of the TSM with a numerical solution of the time
dependent Schr\"odinger equation (TDSE) then demonstrated excellent
quantitative agreement for atomic hydrogen, and significantly
improved agreement for the $\rm  H_2^+$ ion. The argument why using
the acceleration form improves the TSM so much was that the TSM then
becomes correct to first order in the binding potential (we shall
henceforth refer to the TSM obtained this way as $\rm1^{st}$-order
TSM), whereas
 the standard TSM \cite{Corkum, Lewenstein,IvanovBrabec} ($\rm0^{th}$-order TSM) is
 correct to zeroth order in the binding potential.

This work is a followup on Ref.~\cite{GKPRL}, and it has three
goals. First, the first-order accuracy of the $\rm1^{st}$-order TSM
is established. Second, a detailed and general comparison between
the $\rm1^{st}$-order TSM and the $\rm 0^{th}$-order TSM is given.
It is shown that the two always disagree with respect to the
connection between the orbital symmetry and the positions of the
minima in the HHG spectrum, as demonstrated for $\rm H_2^+$
\cite{GKPRL, Bandrauk1}. Third, the opportunity of re-deriving the
SFA is used to give a perspective on some delicate issues in the
derivation, such as gauge and translation invariance, which have
been under debate \cite{Milonni,Kopold,
dispute,BeckerBauerGauge,ChLe06}.

This paper is organized as follows. In Sections \ref{SECSFA} and
\ref{SECZTSM} the derivation of the SFA and $\rm0^{th}$-order TSM is
reviewed, in an attempt to illuminate some subtleties in the
derivation and in order to set the stage for deriving the
$\rm1^{st}$-order TSM in Sec.~\ref{SEC1TSM}. Section \ref{SECOMP}
compares the $\rm0^{th}$-order and $\rm1^{st}$-order TSMs, and
Sec.~\ref{GAUGE} is dedicated to the discussion of gauge and
translation invariance issues in the SFA. Section \ref{DISCUSSION}
gives a brief summary.

Atomic units are adopted throughout the paper.

\section{Strong field approximation} \label{SECSFA}
This work is restricted to the single-electron approximation, where
an atom or a molecule are modeled by an electron in an effective
(local) potential $V(\mathbf r)$:
\begin{equation}\label{H0}
H_0=-\frac{\nabla^2}{2}+V(\mathbf r)+I_p,
\end{equation}
$I_p$ is the binding energy of the ground state, and is added to
\eq{H0} for convenience reasons, such that the ground state of $H_0$
has zero energy. The atom or molecule is placed in a linearly
polarized electric field $E(t)$. The $x$ axis is chosen along the
direction of polarization, the wavelength is assumed sufficiently
long such that the dipole approximation holds, and the length gauge
is chosen, to give the Hamiltonian
\begin{equation}\label{Ht}
H(t)=H_0 - E(t) x.
\end{equation}

The SFA is usually presented [see, e.~g.~Ref.~\cite{KnightReview}
and references therein] as a perturbative expansion in $V(\mathbf
r)$, where the unperturbed Hamiltonian is the Volkov Hamiltonian
\begin{equation}\label{VolkovH}
H_V(t)= -\frac 1 2 \nabla ^2 - E(t) x + I_p.
\end{equation}
The evolution operator $U_V(t,t')$ of $H_V(t)$, defined by
\begin{equation}\label{evoluVolk}
i\partial_t U_V(t,t')=H_V(t)U_V(t,t');\quad U_V(t',t')= \mathbbm{1},
\end{equation}
is known exactly \cite{Lewenstein}. Using the Lippmann-Schwinger
 equation one can approximate $U(t,t')$, the evolution
operator associated with $H(t)$ (defined through \eq{evoluVolk} with
all $V$ subscripts omitted) to arbitrary order in $V(\mathbf r)$
\cite{Milo1}. The zeroth and first order would be
\begin{eqnarray}\label{PT1}
U_0(t,t')&=&U_V(t,t')\\
\label{PT2} U_1(t,t')&=&-i\int_{t'}^t dt'' U_V(t,t'')VU_V(t'',t').
\end{eqnarray}
With the operator $U(t,t')$ at hand, any problem can be solved.

If at time $t=0$ the electron is in the ground state $|0\rangle$ of
$H_0$, the wavefunction $|\psi(t)\rangle$ at any time would be given
by
\begin{equation}\label{exacteq}
 |\psi(t)\rangle=U(t,0)|0\rangle.
\end{equation}
One therefore could suggest approximating $U(t,t')$ in \eq{exacteq}
by the perturbative expansion outlined in Eqs.~(\ref{PT1},
\ref{PT2}). However this is not the way the SFA is usually
performed. The route is rather by making the ansatz
\cite{Lewenstein}
\begin{equation}\label{ansatz}
|\psi\rangle=a(t)|0\rangle + |\varphi(t)\rangle.
\end{equation}
$a(t)$ has the initial conditions $a(0)=1$ and is determined later.
In other words, the way we split $|\psi\rangle$ into $a(t)|0\rangle$
and $|\varphi(t)\rangle$ is not yet defined.

Let now  $|\varphi(t)\rangle$ be the exact solution of
\begin{equation}\label{phieq1}
i|\dot\varphi(t)\rangle=H(t)|\varphi(t)\rangle-a(t) E(t) x|0\rangle,
\end{equation}
which in terms of $U$ reads
\begin{equation}\label{exactprop} |\varphi(t)\rangle=-i\int_0^t
U(t,t')E(t')xa(t')|0\rangle dt'.
\end{equation}
Let us assume for the moment that $a(t)$ is specified. The SFA
approximates $U$ in \eq{exactprop} by a perturbation series in $V$
upon $U_V$.

One can now ask why the SFA approximates $U$ in \eq{exactprop}
rather than in \eq{exacteq}, if both lead to an exact wavefunction
when $U$ is exact? In other words, why one needs the ansatz
(\ref{ansatz}) and the rather non-straightforward \eq{phieq1}? To
our understanding, the answer to these questions is that the
perturbed $U_V(t,t')$ is a very bad approximation for describing the
evolution of (quasi) bound states.

Without the laser field, perturbation theory for the evolution
operator (the Born series) diverges at bound states of the full
Hamiltonian \cite{Newton}. The presence of a laser field may change
the situation, since formally there are no bound states at all. We
are not in a position to make a rigorous mathematical statement
about the convergence of the SFA, but since the ground state remains
quasi-bound, one may expect convergence difficulties. Obviously, at
least for the the zeroth order theory, $U_V(t,t')$ itself gives a
very bad approximation for the evolution of a quasi-bound state. In
fact, in the Keldysh ionization rate \cite{Keldysh} is found rather
indirectly, by calculating the rate at which the norm of the
continuum increases, and the rate at which the ground-state decays
is inferred solely by conservation of the total norm.

The last term in \eq{phieq1} is therefore added because we know in
advance that we are going to use an approximate propagator, which
will give a very bad description for the evolution of the ground
state. Yet, \eq{phieq1} provides a clear route for systematically
improving the SFA: $a(t)$ is first assumed to be known, and
\eq{phieq1} is solved to in principle an arbitrary order in
$V(\mathbf r)$. Then $a(t)$ is found by demanding the conservation
of norm, or by means of other approximations, as the perturbed
$U_V(t,t')$ alone is not well suited for this purpose. Note that in
the limit where $U$ is exact, one easily finds from \eq{phieq1} and
\eq{ansatz} that $a(t)\equiv1$.

\section{Zeroth order TSM} \label{SECZTSM}
In this section we re-derive the standard $\rm0^{th}$-order TSM. The
main reason is that our slightly different derivation lays the
technical foundations for establishing the first order accuracy of
the $\rm1^{st}$-order TSM in Sec.~\ref{SEC1TSM}.

The $\rm0^{th}$-order TSM is obtained by simply replacing $U(t,t')$
by $U_V(t,t')$ in \eq{phieq1}. Since $U_V(t,t')$ is known exactly,
the solution can be written in a closed form \cite{Lewenstein}:
\begin{eqnarray}\label{psip}
\shoveleft {\langle {\bf p-A}(t)|
\varphi(t)\rangle}=\phantom{AAAAAAAAAAAAAAA}\cr = -i\int\limits_0^t
dt' a(t')E(t')\langle {\bf p}-{\bf A}(t')|x |0\rangle e^{-i S({\bf
p},t,t')}.
\end{eqnarray}
${\bf A}(t)\equiv -{\bf \hat x}\int^t E(t')dt'$ is a vector
potential that describes the electric field $E(t)$, and
\begin{equation}\label{S}
S({\bf p},t,t')\equiv\frac 1 2 \int\limits _{t'}^t  ({\bf p}-{\bf
A}(t''))^2dt'' +I_p(t-t').
\end{equation}

At this point we slightly deviate from the standard derivation of
the TSM \cite{Lewenstein, IvanovBrabec}: We perform the saddle-point
integration in \eq{psip} right now, before proceeding. This
calculation was pioneered by Keldysh \cite{Keldysh} and is further
discussed in a vast number of SFA studies [see
e.~g.~Ref.~\cite{DeloneKrainov} and references therein]. The
discussion in this work is limited to the tunneling regime, which is
defined by the requirements \cite{DeloneKrainov}:
\begin{subequations}\label{allreqs}
\begin{eqnarray}
\label{reqs} E_0&\ll& (2I_p)^{3/2}\\ \label{reqs1}
\gamma\equiv\frac{\omega}{E_0}\sqrt{2I_p}&\ll&1 \\
\label{reqs2} \omega&\ll& I_p
\end{eqnarray}
\end{subequations}
In \eq{reqs} $E_0$ is the amplitude of the driving field and
$\omega$ is its frequency. These parameters are precisely defined
for a sinusoidal driving field, and more loosely for a general
field, such as a sinusoidal field with an envelope. In the latter
case, $\omega$ can be thought of as the parameter characterizing the
timescale over which $E(t)$ varies. $\gamma$ is the well known
Keldysh parameter \cite{Keldysh}. Note that the requirement
(\ref{reqs2}) follows from (\ref{reqs}) and (\ref{reqs1}).

Under the conditions (\ref{allreqs}), the saddle-point integration
of \eq{psip} is carried out to give
\begin{eqnarray}\label{Keld111}
{\langle {\bf p-A}(t)| \varphi(t)\rangle}=
\frac{\sqrt[4]{2I_p}}{\sqrt\pi}\sum _n
a(t_n(p_x))\frac{w((E(t_n(p_x))}{|E(t_n(p_x))|}\times\cr\times
e^{\frac{-p_\perp^2}{|E(t_n(p_x))|\sqrt{2I_p}}} e^{-iS(\mathbf
p,t,t_n(p_x))}.\phantom{A}
\end{eqnarray}
$w(E)$ is the static Stark ionization rate associated with the
ground state, $p_\perp^2\equiv p_y^2+p_z^2$, and the function
$t_n(p_x)$ is the set of positive real solutions of the equation
\begin{equation}\label{sad_real}
p_x=A_x(t_n(p_x)).
\end{equation}
The number of solutions depends on $E(t)$. $a(t)$ can now be found
by requiring conservation of the norm in \eq{ansatz} [neglecting
$\langle 0 | \varphi(t)\rangle$] to give the well-known expression
\cite{BrabecReview}
\begin{equation}\label{aa}
|a(t)|^2=e^{-\int_0^t w(E(t'))dt'}.
\end{equation}

Using the ansatz (\ref{ansatz}) with the wavefunction
(\ref{Keld111}), one can now compute the expectation value of the
dipole moment:
\begin{eqnarray}\label{dip}
\langle x \rangle &=& |a(t)|^2\langle 0 |x|0\rangle+\langle
\varphi(t) | x | \varphi(t)\rangle+\xi_0(t)+\xi_0^*(t),\phantom{AA}
\end{eqnarray}
where
\begin{equation}\label{xi0}
\xi_0(t)\equiv a^*(t)\langle 0 | x |\varphi(t)\rangle.
\end{equation}
The origin of coordinates can always be chosen such that the first
term in \eq{dip} vanishes. The second term has no high harmonics
\cite{IvanovRza}, since matrix elements of $x$ between two different
Volkov states vanish. The high harmonics come from the cross term
$\xi_0(t)$. Using \eq{Keld111} we find:
\begin{eqnarray}\label{xip}
\xi_0(t)=a^*(t)\frac{\sqrt[4]{2I_p}}{\sqrt\pi}\sum _n\int d^3p
\langle 0 |x| {\bf p-A}(t)\rangle a(t_n(p_x)) \times\cr\times
 \frac{w((E(t_n(p_x))}{|E(t_n(p_x))|}e^{\frac{-p_\perp^2}{|E(t_n(p_x))|\sqrt{2I_p}}} e^{-iS(\mathbf
 p,t,t_n(p_x))}\phantom{A}
\end{eqnarray}
The $\mathbf p$ integration is now carried out in the stationary
phase approximation as in Ref.~\cite{Lewenstein}. The stationary
phase is attained at $p_\perp=0$ and at all $p_x$ values such that
$t_n(p_x)$ satisfy [see Appendix \ref{tunnel} for a more careful
discussion]
\begin{equation}\label{statp}
\int_{t_n}^t(A(t_n)-A(t''))dt''=0.
\end{equation}

For any given $t$, $\bar t_n(t)$ is defined as the set of solutions
to \eq{statp}, i.~e.~birth times of trajectories that end up at the
origin at time $t$. Then in the stationary phase approximation
\eq{xip} gives
\begin{eqnarray}\label{lewedip2}
\xi_0(t)&=& e^{\frac{-i\pi}{4}}\sqrt[4]{2I_p}2^{3/2}\pi\sum_n
\frac{e^{-i \overline S_n(t)}}{(t-\overline t_n(t))^{3/2}} a^*(t)
\times \cr &\times& a(\overline t_n(t)) \langle 0|x | \mathbf
A(\overline t_n(t))-\mathbf A( t)\rangle \frac{w((E(\bar
t_n(t))}{|E(\bar t_n(t))|}, \phantom{AA}
\end{eqnarray}
where $\bar S_n(t)\equiv S(\mathbf A(\bar t_n(t)),t,\bar t_n(t))$.

\eq{lewedip2} is identical to the original TSM expression derived in
Ref.~\cite{IvanovBrabec}. The only difference is that \eq{lewedip2}
takes into account the depletion of the ground state, and that all
numerical prefactors ($2^{ 3/2}\pi$) were calculated. The
re-derivation of the $\rm0^{th}$-order TSM is now concluded.

\section{First order TSM}\label{SEC1TSM}
A straight forward way to upgrade the TSM from zeroth to first order
accuracy in $V$ would be improving the TSM wavefunction (\ref{psip})
by adding the first-order correction to the evolution operator,
\eq{PT2}. This method has been employed in Ref.~\cite{Lew95}, for
improving the SFA theoretical description of above threshold
ionization. For HHG we propose another method, which is
significantly simpler, and does not require correcting the
wavefunction.

If $|\psi(t)\rangle$ is the exact solution of the time-dependent
Schr\"odinger equation with the Hamiltonian (\ref{Ht}), then the
Ehrenfest theorem holds:
\begin{subequations}\label{Ehren1}
\begin{eqnarray}\label{dlength}
&{\phantom=}&\frac{d^2\langle\psi(t)| x|\psi(t) \rangle}{dt^2}=\\
\label{dvelocity}&=& \frac{d\langle\psi(t)| p_x|
\psi(t)\rangle}{dt}=\\ \label{daccel} &=&-\langle \psi(t)|
\partial_x V(\mathbf r)|\psi(t) \rangle +E(t)
\end{eqnarray}
\end{subequations} Therefore computing the time-dependent dipole
expectation value in all three forms of \eq{Ehren1}, length
(\ref{dlength}), velocity (\ref{dvelocity}), or acceleration
(\ref{daccel}), is equivalent if accompanied by the appropriate
differentiation or integration in time.

However with the approximate wavefunction given by
Eqs.~(\ref{ansatz}) and (\ref{Keld111}), the results will be in
general different for each form. One therefore has to make a choice
which form to use for this particular wavefunction. We argue that
the acceleration form gives in general the best results, since even
if $|\varphi(t)\rangle$ is only correct to zeroth order in
$V(\mathbf r)$, the expectation value is automatically correct, at
least formally, to first order in $V(\mathbf r)$. One therefore
needs to evaluate
\begin{eqnarray}\label{dipEh}
\frac{d^2\langle x \rangle}{dt^2}
 &=& \ddot \xi_1(t)+\ddot \xi^*_1(t)+\ddot\xi_c(t)
\end{eqnarray}
where
\begin{eqnarray}\label{xi1}
 \ddot \xi_1(t)&\equiv& -a^*(t)\langle 0 |
\partial_x V(\mathbf r) |\varphi(t)\rangle \\ \label{xic}
\ddot\xi_c(t)&\equiv&-\langle \varphi(t) | \partial_x V(\mathbf r) |
\varphi(t)\rangle
\end{eqnarray}
The last term of \eq{daccel} has been dropped, since it is the
driving field itself and thus contains no harmonics. It is easy to
show, exploiting the commutator  $[p,H_0]$, that $\langle 0
|\partial_x V(\mathbf r)|0\rangle=0$ [irrespectively of the position
of origin], which is why the corresponding term in \eq{dipEh} is
missing.

Note that the mere appearance of $V$ in \eq{dipEh} is not sufficient
to warrant first order accuracy. One can see that, for example, from
the fact that zeroth order SFA expressions in $V$ can be transformed
such that they appear linear or quadratic in $V$ \cite{Lohr,
BeckerLew}. The (formal) first-order accuracy of \eq{dipEh} is
established by the fact that a first-order correction in $V$ to the
wavefunction in \eq{psip} will result in only a second order
correction in $V$ in \eq{dipEh}. This is why we dedicated
Sec.~\ref{SECSFA} to carefully defining the procedure by which the
SFA is corrected order by order in $V$.

The cross term $\ddot \xi_1$ has the same structure as $\xi_0$
[\eq{xi0}]. Going through the same steps that have led to
\eq{lewedip2}, one arrives at the same expression $\ddot \xi_1$,
with the only difference that the matrix element $\langle 0
|x|\mathbf k\rangle$ is replaced by $-\langle 0 |\partial_xV(\mathbf
r)|\mathbf k\rangle$:
\begin{eqnarray}\label{lewedip3}
\ddot \xi_1(t)&=&- e^{\frac{-i\pi}{4}}\sqrt[4]{2I_p}2^{3/2}\pi
\sum_n \frac{e^{-i \overline S_n(t)}}{(t-\overline t_n(t))^{3/2}}
\times \cr &\times& a^*(t)a(\overline t_n(t)) \frac{w((E(\bar
t_n(t))}{|E(\bar t_n(t))|}\cr &\times&\langle 0|\partial_xV(\mathbf
r) | \mathbf A(\overline t_n(t))-\mathbf A( t)\rangle. \phantom{AA}
\end{eqnarray}
\eq{lewedip3} is the improved version of the TSM presented in
Ref.~\cite{GKPRL}. The change in the expression compared to
\eq{lewedip2} is very small and easy to implement, and yet results
in a very large difference, especially in the case of molecules.

\eq{lewedip3} is an expression for $\ddot\xi_1$, but \eq{dipEh}
contains also $\xi_c$. It turns out that the contribution of $\ddot
\xi_c$ to HHG is smaller by at least $O(\omega^{3/2})$ than that of
$\ddot\xi_1$. $\ddot \xi_c$ is therefore negligible for
$\omega\ll1$, which coincides with \eq{reqs2} when $I_p$ is of
$O(1)$. The latter holds for all neutral (or not highly charged)
atoms and molecules. The evaluation of $\ddot \xi_c$ is rather
lengthy, especially for potentials with a long-ranged Coulomb tail,
and is given in Appendix \ref{CC}.

\section{$\bf0^{th}$ vs. $\bf 1^{st}$-order TSM  -- comparison}\label{SECOMP}

The $\rm0^{th}$-order TSM suggests \eq{xi0} as an approximation to
the dipole moment, whereas the $\rm1^{st}$-order TSM suggests
\eq{xi1}. In order to compare them conveniently, we now
differentiate \eq{xi0} twice in time and compare $\ddot \xi_0$ with
$\ddot \xi_1$. This is done in detail in Appendix \ref{diffxhi0}.
The result is that under the condition (\ref{reqs}), in order to
obtain $\ddot \xi_0(t)$, one has to replace the $\rm1^{st}$-order
TSM recombination amplitude
\begin{equation}\label{arecnew}
a_{\rm rec}^{\rm new} (\mathbf k)=\langle 0|\partial_x V(\mathbf
r)|\mathbf k\rangle,
\end{equation}
in \eq{xi1} by
\begin{equation}\label{arecold}
a_{\rm rec}^{\rm old}(\mathbf k)=(I_p+\tfrac 1 2|\mathbf
k|^2)^2\langle 0|x|\mathbf k\rangle.
\end{equation}
Comparing the $\rm1^{st}$-order TSM and the $\rm0^{th}$-order TSM is
thus reduced to comparing $a_{\rm rec}^{\rm new}$ and $a_{\rm
rec}^{\rm old}$ respectively.

The two expressions look different, and indeed they are. A detailed
comparison requires the knowledge of $V$. However in order to gain
some general insight, in what follows we study the $k\to\infty$
asymptotic behavior of $a_{\rm rec}^{\rm new}$ and $a_{\rm rec}^{\rm
old}$. Some pretty general statements can be made about
$k\to\infty$, which turn out to provide important insights also for
$k$-s of $O(1)$.

\subsection{High-momentum asymptotic evaluation}
$a_{\rm rec}^{\rm old}(\mathbf k)$ can be written as
\begin{equation}\label{FT}
a_{\rm rec}^{\rm old}(\mathbf k)=\frac{1}{(2\pi)^{3/2}}\int
\chi^{\rm old}(\mathbf r)e^{-i \mathbf k \cdot \mathbf r}dr
\end{equation}
(and similarly for $a_{\rm rec}^{\rm new}(\mathbf k)$ and $\chi^{\rm
new}(\mathbf r)$), where
\begin{eqnarray}\label{chidef}
\chi^{\rm old}(\mathbf r)&\equiv&(I_p-\tfrac 1 2
\nabla^2)^2x\psi_0(\mathbf r)\\ \label{chidef2} \chi^{\rm
new}(\mathbf r)&\equiv&\psi_0(\mathbf r)\partial_x V(\mathbf r),
\end{eqnarray}
with $\psi_0(\mathbf r)\equiv \langle \mathbf r | 0 \rangle$. Using
$H_0|0\rangle=0$, \eq{chidef} can be transformed into
\begin{eqnarray}\label{chi1}
\chi^{\rm old}(\mathbf r)&=&2\psi_0(\mathbf r)\partial_x V(\mathbf
r)+2V(\mathbf r)\partial_x\psi_0(\mathbf r)\cr &+&x V^2(\mathbf
r)\psi_0(\mathbf r)+x\nabla V(\mathbf r)\cdot \nabla \psi_0(\mathbf
r)+\cr &+&\tfrac 1 2 x\psi_0(\mathbf r)\nabla^2V(\mathbf r)
\end{eqnarray}
The $k\to\infty$ asymptotic behavior of $a_{\rm rec}^{\rm
old}(\mathbf k)$ and $a_{\rm rec}^{\rm new}(\mathbf k)$ is dictated
by the most singular part $\chi^{\rm old}(\mathbf r)$ and $\chi^{\rm
new}(\mathbf r)$ respectively, since they are connected by a Fourier
transform \cite{Simon_Reed}.

We now assume that $V(\mathbf r)$ is an effective potential that
represents a molecule (or, as a special case, an atom) with nuclei
at positions $\mathbf r_j$ with charges $Z_j$, $j=1...N$. Common
choices for an effective potential to model such a system have a
$Z_j/|\mathbf r-\mathbf r_j|$ singularity near the $j$-th nucleus,
which is partially screened off away from the nucleus. Then the last
term in \eq{chi1} can be written as
\begin{equation}\label{deltas}
\tfrac 1 2 x\psi_0(\mathbf r)\nabla^2V(\mathbf r)=-2\pi  x
\psi_0(\mathbf r)\sum_j Z_j\delta(\mathbf r-\mathbf r_j)
\end{equation}
Now one has to distinguish two cases. The first case is a molecule.
Whenever there is more than one nucleus, \eq{deltas} does not vanish
[apart from possibly a very few orientations of the molecule]. Then
the term in \eq{deltas} is the most singular one among all terms on
the right hand side of \eq{chi1}, and thus governs $a_{\rm rec}^{\rm
old}(\mathbf k)$ at $ k \to \infty$. In this case
\begin{eqnarray}\label{chi2}
\chi^{\rm old}(\mathbf r)\sim -2\pi\sum_j  x_j \psi_0(\mathbf r_j)
Z_j\delta(\mathbf r-\mathbf r_j)
\end{eqnarray}
where the notation `$\sim$' means `equal in its most singular part'.

The second case is an atom, which has only one nucleus, whose
position can be always chosen at $\mathbf r=0$ [see Sec.~\ref{GAUGE}
for a discussion about translation invariance]. In this case
\eq{deltas} obviously vanishes. Then the most singular term on the
right hand side of \eq{chi1} is the first one, and we find that
\begin{eqnarray}\label{chi3}
\chi^{\rm old}(\mathbf r)\sim 2 \chi^{\rm new}(\mathbf r)
\end{eqnarray}

\subsection{Atoms}
\eq{chi3} implies that for atoms
\begin{equation}\label{atoms}
a_{\rm rec}^{\rm old}(\mathbf k)\to 2a_{\rm rec}^{\rm new}(\mathbf
k),
\end{equation}
where `$\to$' denotes `approaches asymptotically at $k\to\infty$'.
Thus $a_{\rm rec}^{\rm new}(\mathbf k)$ and $a_{\rm rec}^{\rm
old}(\mathbf k)$ do not agree even asymptotically for $k\to\infty$,
the limit were plane waves approximate the exact continuum states
increasingly well. For atoms,  the $\rm0^{th}$-order TSM predicts a
4 times larger HHG yield than the ($\rm1^{st}$-order TSM). For a
$Z/r$ singularity in the potential, the asymptotic behavior of
$a_{\rm rec}^{\rm new}$ can be easily worked out to give
\begin{equation}\label{sing}
a_{\rm rec}^{\rm new}(\mathbf k)\to \ii Z\psi_0(0)\sqrt \frac{2 }{
\pi}\frac{k_x}{k^2}
\end{equation}
When $\psi_0(0)$ vanishes \eq{sing} is modified.

\begin{figure}[htb]
\centering
\includegraphics[width=9cm]{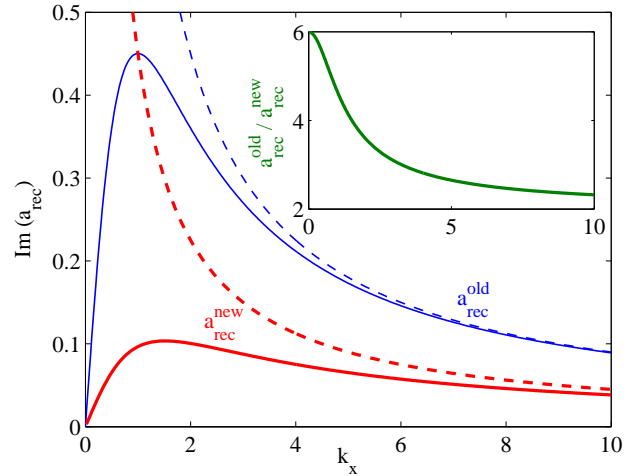}
\caption{comparison of $a_{\rm rec}^{\rm new}$ [thick red lines] and
$a_{\rm rec}^{\rm old}$ [thin blue lines] for hydrogen. The
continuous lines correspond to the two expressions in \eq{rec_coul}.
The dashed lines are the asymptotic expressions [\eq{atoms} and
\eq{sing}]. The inset shows the ratio between the two expressions in
\eq{rec_coul}. }\label{fig1}
\end{figure}

For finite $k$ the disagreement between $a_{\rm rec}^{\rm
new}(\mathbf k)$ and $a_{\rm rec}^{\rm old}(\mathbf k)$ can be even
greater. For the hydrogen atom one finds \cite{GSKPRA}
\begin{eqnarray}\label{rec_coul}
a_{\rm rec}^{\rm new}(k_x)&=&\frac{i\sqrt 2}{\pi}
\frac{k_x-\tan^{-1}k_x}{k_x^2} \cr a_{\rm rec}^{\rm
old}(k_x)&=&\frac{i 2^{3/2}}{\pi}\frac{k_x}{1+k_x^2}
\end{eqnarray}
for $\mathbf k$ lying along the $x$ axis. Fig.~\ref{fig1} compares
the two expressions in \eq{rec_coul}.

\subsection{Molecules}
The differences between the $\rm0^{th}$-order TSM and the
$\rm1^{st}$-order TSM for molecules is particularly striking. The
two models give \emph{opposite} relations between orbital symmetry
and the positions of the minima in the HHG spectrum, a subject of
many recent theoretical and experimental studies \cite{Lin,
Marangos1, Kanai, Chang, Nalda, Lein2002, Knight}. A set of minima
is predicted by the $\rm0^{th}$-order TSM with an odd orbital,
corresponds to an even orbital in the $\rm1^{st}$-order TSM, and
vice versa.

From \eq{chi2} we find that for molecules
\begin{eqnarray}\label{chi22}
a_{\rm rec}^{\rm old}(\mathbf k)\to-\frac{1}{\sqrt{2\pi}}\sum_j  x_j
\psi_0(\mathbf r_j) Z_je^{-\ii \mathbf k\cdot \mathbf r_j},
\end{eqnarray}
whereas from \eq{arecnew} and \eq{sing} we find
\begin{eqnarray}\label{chi33}
a_{\rm rec}^{\rm new}(\mathbf k)\to\ii \sqrt \frac{2 }{
\pi}\frac{k_x}{k^2}\sum_j \psi_0(\mathbf r_j) Z_je^{-\ii \mathbf
k\cdot \mathbf r_j},
\end{eqnarray}
Observing \eq{chi22} and \eq{chi33}, it becomes clear why the
$\rm0^{th}$-order TSM and $\rm1^{st}$-order TSM disagree about the
connection between the symmetry $\psi_0$ and the zeros (or minima)
of $|a_{\rm rec}|$. Apart from the additional $k_x/k^2$ envelope in
\eq{chi33}, which does not affect the position of the zeros,
\eq{chi33} is the Fourier transform of $x\psi_0(\mathbf r)$, whereas
\eq{chi22} is the Fourier transform of $\psi_0(\mathbf r)$ (both
sampled at the singular points). $x\psi_0(\mathbf r)$ and
$\psi_0(\mathbf r)$ obviously have the opposite symmetry with
respect to a $x\to-x$ reflection.

In order to illustrate this we consider the $\rm H_2^+$ ion, where
the nuclei are positioned along the $x$ axis, at $x=\pm R/2$. In
this case \eq{chi22} gives
\begin{eqnarray}\label{chi222}
a_{\rm rec}^{\rm old}(\mathbf k)\to\frac{\ii RC}{\sqrt{2\pi}}
\sin(\tfrac12 k_x R),
\end{eqnarray}
where $C\equiv \psi_0(-\frac 1 2 R\mathbf {\hat x})=\psi_0(\frac 1 2
R\mathbf {\hat x})$. In contrast, \eq{chi33} gives
\begin{eqnarray}\label{chi333}
a_{\rm rec}^{\rm new}(\mathbf k)\to\ii \sqrt \frac{2 }{
\pi}\frac{k_x}{k^2} 2C \cos(\tfrac 1 2 k_x R),
\end{eqnarray}

A zero of $a_{\rm rec}$ at a given $k_x$ corresponds to a minimum in
the HHG spectral intensity at the frequency $I_p+\frac 1 2 k_x^2$.
The $\rm0^{th}$-order TSM therefore predicts minima at energies
given by $I_p+(2\pi n/R)^2$ ($n$ is an integer), whereas the
$\rm1^{st}$-order TSM predicts minima at energies given by
$I_p+(2\pi (n+\frac 1 2)/R)^2$. The latter condition well agrees
with numerically-exact results \cite{Knight, GKPRL}.

\begin{figure}[htb]
\includegraphics[width=9cm]{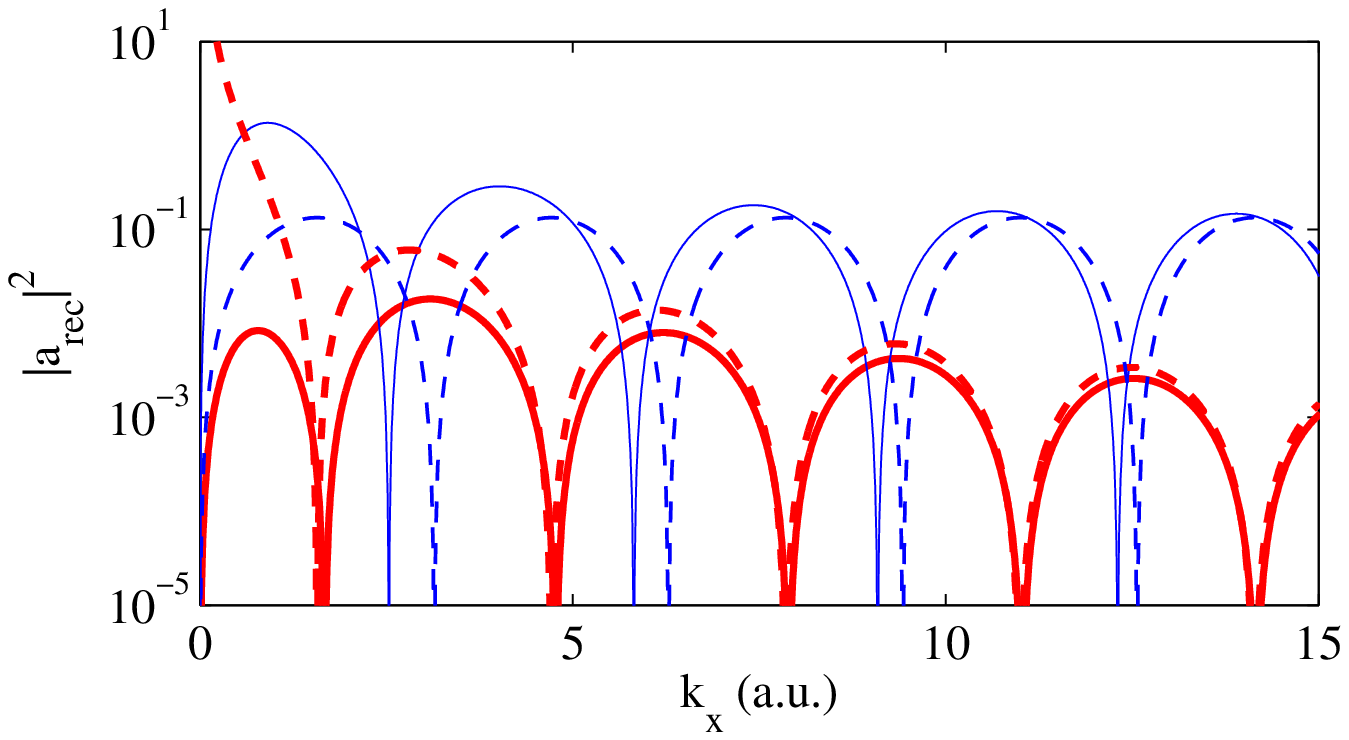}
\caption{$|a_{\rm rec}^{\rm old}|^2$ (thin blue line) and $|a_{\rm
rec}^{\rm new}|^2$ (thick red line) as function of $v_x$ for $\rm
H_2^+$ (R=2). The dashed lines are the corresponding asymptotic
($v_x\gg1$) approximations [\eq{chi222} and
\eq{chi333}].}\label{fig3}
\end{figure}

It is easy to see from \eq{chi22} and \eq{chi33} that if
$\psi_0(\mathbf r)$ were an odd rather than even wavefunction with
respect to $x\to -x$, the cosine [sine] in \eq{chi222} [\eq{chi333}]
is replaced by a sine [cosine]. Therefore if the $\rm0^{th}$-order
TSM is used for reconstructing $\psi_0(\mathbf r)$  from the HHG
spectrum, it flips the symmetry of $\psi_0(\mathbf r)$ from even to
odd and vice versa. This statement is, of course, based only on the
$k_x\to\infty$ asymptotic behavior of the $a_{\rm rec}$-s. Yet it is
interesting to note that for $\rm H_2^+$, \eq{chi222} and
\eq{chi333} approximate \eq{arecnew} and \eq{arecold} reasonably
well for low momenta, as Fig.~\ref{fig3} shows.

Note that the $k_x/k^2$ envelope, which is present in \eq{chi33} but
absent in \eq{chi22}, results in an orders-of-magnitude difference
in the magnitudes of the $a_{\rm rec}$-s, in addition to the
distorted shape.

\section{Gauge and translation invariance issues} \label{GAUGE}

The dependence of the SFA on gauge has been the subject of lively
discussions \cite{Reiss,Milonni, dispute, BeckerBauerGauge,
Kopold,ChLe06}. Here we do not attempt a comprehensive discussion of
this subject. However since some steps in Sec.~\ref{SECOMP} seem to
rely on placing the origin of the laser potential and the atom at
the same point, we briefly address the related issue of translation
invariance.

Consider thus the Hamiltonian
\begin{equation}\label{Htgauge} H(t)=-\frac 1 2 \nabla
^2 +V(\mathbf r) - E(t) (x-x_0)+I_p
\end{equation}
($x_0$ is a constant), which is related to \eq{Ht} by a gauge
transformation. The exact solution of the TDSE with \eq{Htgauge}
obviously gives the same time-dependent expectation values as with
\eq{Ht}. However the SFA can give different results for \eq{Htgauge}
and \eq{Ht}. In particular, the SFA with \eq{Htgauge} can give rise
to the generation of even harmonics \cite{Kopold}, which is an
artifact.

There is an obvious way by which this problem can be cured, namely,
by replacing the ansatz (\ref{ansatz}) by
\begin{equation}\label{ansatzGauga}
|\psi\rangle=a(t)e^{-ix_0\int E(t)dt}|0\rangle + |\varphi(t)\rangle.
\end{equation}
Substituted \eq{ansatzGauga} into the TDSE with the Hamiltonian
(\ref{Htgauge}), it is easy to convince oneself that the results no
longer depend on $x_0$. This is because wavefunction (\ref{ansatz})
undergoes exactly the gauge transformation that connects
\eq{Htgauge} to \eq{Ht}, guaranteeing gauge invariance. The SFA
depends on the choice of gauge simply because it crucially depends
on the choice of the initial ansatz.

One can now ask how to obtain an ansatz for a given gauge a-priori,
without e.~g.~gauge-transforming \eq{ansatz}, and how to obtain the
ansatz (\ref{ansatz}) itself a-priori. Our answer is that the ansatz
should provide the best approximation for the evolution of the
quasi-bound ground state, since the (perturbed) Volkov evolution
operator cannot do it [see Sec.~\ref{SECSFA}].

In the low frequency regime [\eq{reqs2}], a good approximation for
the evolution of the ground state can be found using the adiabatic
theorem \cite{adiabatic}: It is approximately given by $e^{-i\int
\varepsilon(E(t))dt}|\phi(E)\rangle$, where the $|\phi(E)\rangle$
and $\varepsilon (E)$ are the field-dependent ground state
wavefunction and energy, defined through the eigenvalue equation
\begin{equation}\label{eigen}
\left(-\tfrac 1 2 \nabla ^2 +V(\mathbf r) - E
(x-x_0)+I_p\right)|\phi(E)\rangle=\varepsilon(E)|\phi(E)\rangle.
\end{equation}
The adiabatic theorem holds even though $|\phi(E)\rangle$ becomes a
resonance \cite{NimrodAdiabatic}. If we neglect the Stark shift of
the energy, we obtain $\varepsilon(E)=Ex_0$. Going further and
approximating $|\phi(E)\rangle$ by $|\phi(0)\rangle=|0\rangle$, we
obtain exactly the ansatz (\ref{ansatzGauga}), a-priori, without
referring to \eq{ansatz}. In fact, \eq{ansatz} itself is obtained
for $x_0=0$.

The discussion can be extended to at least one common gauge, namely
the velocity gauge. Then the adiabatic argument would again give a
modified ansatz, similar to the one used in
Ref.~\cite{BrabecDipole}. It is likely that the differences between
the length and velocity gauges reported in
Ref.~\cite{BeckerBauerGauge} would then disappear.

\section{Discussion}  \label{DISCUSSION}

The large discrepancy between the $\rm0^{th}$-order TSM HHG spectra
and numerically-exact calculations has been attributed long ago
\cite{BrabecReview} to the SFA wavefunction being inaccurate. It is
often argued that this SFA wavefunction is especially inaccurate for
molecules \cite{Greene}. This work strongly supports these ideas:
The length and acceleration forms of the dipole operator used on the
same SFA wavefunction are shown to give dramatically different
results, especially for molecules. For an exact wavefunction the
results would be identical, and thus the large discrepancy is
evidence for the inaccuracy of the wavefunction.

Since the wavefunction is so far from being exact, it is especially
important to use it properly. In this work we have argued that
together with the acceleration form of the dipole, the zeroth-order
SFA wavefunction leads to an approximation (the $\rm1^{st}$-order
TSM) with first-order overall accuracy in the binding potential.
Since the wavefunction itself is so inaccurate, it can lead to very
large errors if used otherwise than in the specially-suited way
introduced in this work. The accuracy of the $\rm1^{st}$-order TSM
should be tested in terms of the expectation values it generates and
not in terms of the wavefunction itself.

We have shown that the continuum-continuum term in the
$\rm1^{st}$-order TSM is negligible compared to the bound-continuum
term. Therefore, the widely-used $\rm0^{th}$-order TSM is upgraded
to first-order accuracy by simply replacing $x$ by $\partial_x
V(\mathbf r)$ in the recombination amplitude, which is simple to
implement. The $\rm1^{st}$-order TSM shows excellent agreement with
numerically-exact results for atoms \cite{GKPRL, SG06} and good
agreement for $\rm H_2^+$ \cite{GKPRL}.

The SFA decomposes the wavefunction into a continuum, and a
quasi-bound ground state. This is justified, since the strong laser
field smears the excited states of the unperturbed system and turns
them into a continuum \cite{Lewenstein}. The dynamics of the
continuum is approximated by the Volkov propagator, which can be
corrected order by order in the binding potential. The quasi-bound
state however should be treated separately.

It should be noted that there could be more than one quasi-bound
state. For example, the initial state of the electron can be an
excited state of the laser-free Hamiltonian $H_0$, a case that was
not treated in this work. This is commonly the case in
single-electron models of multi-electron atoms. Then the states
lying energetically below the initial state are also quasi-bound,
and should also be treated separately. One simple way to do that is
to project these states out of the SFA wavefunction, as briefly
discussed at the end of Ref.~\cite{SG06}.

\acknowledgments

The authors thank M. Yu. Ivanov and R. Santra for fruitful
discussions. Support by DARPA under contract FA9550-06-1-0468 is
gratefully acknowledged.

\appendix

\section{Complementary remarks on the derivation of the $\rm0^{th}$-order TSM}
\label{tunnel} \eq{statp} is gives the stationary phase condition
only to leading order in $\gamma$. In fact, when $S(\mathbf
p,t,t_n(p_x))$ is differentiated with respect $p_x$ taking
\eq{sad_real} into account, the result is
\begin{equation}\label{statp1}
\frac{I_p}{E(t_n)}+\int_{t_n}^t(A(t_n)-A(t''))dt''=0.
\end{equation}
The second term in \eq{statp1} is of the order of $E_0/\omega^2$,
while the first one is of the order of $I_p/E_0$. Therefore the
first term in \eq{statp1} is a second-order correction in $\gamma$
to \eq{statp}. It is interesting to note however that the subleading
term term is meaningful, since the electron is released by tunneling
roughly at the turning point ($x=I_p/E$) \cite{DeloneKrainov} rather
than at the origin.

Note that the denominator $(t-\overline t_n(t))^{3/2}$ in
\eq{lewedip2} cannot result in a divergence. If the first term in
\eq{statp} is retained, the reason is obvious -- the duration of a
trajectory is never zero, since it begins and ends at different
locations. If the first term in \eq{statp} is neglected, as often
happens, one can easily show that trajectories for which $t$
approaches $t_n(t)$ can only occur at points where $E(t)=0$, and
then $w(E)$ vanishes exponentially. Performing the $t'$ integration
first thus spares the need for the regularizing parameter $\epsilon$
used in, e.~g.~Ref.~\cite{Lewenstein}.

Note also that the denominator $(t-\overline t_n(t))^{3/2}$ is
correct only to zeroth order in $\gamma$. The leading correction is
obtained if the $e^{-p_\perp^2/(|E(t_n(p_x))|\sqrt{2I_p}}$ factor in
\eq{xip} is included in the Gaussian in integration and if
\eq{sad_real} is taken into account when the second derivative of
$S$ with respect to $p_x$ is computed.

\section{Evaluation of the continuum-continuum term} \label{CC}
For completeness, in  this Appendix the continuum-continuum term
$\ddot\xi_c(t)$ [\eq{xic}] is evaluated. To this aim we define
\begin{equation}\label{Vdef}
Q(\mathbf k)\equiv \frac{1}{(2\pi)^{3/2}}\int d^3r [\partial_x
V(\mathbf r)] e^{-i \mathbf k \cdot \mathbf r},
\end{equation}
the Fourier transform of the force field derived from the potential
$V$. It turns out that a long-range Coulomb behavior of $V$ requires
more careful evaluation of $\ddot\xi_c$ than a short-ranged $V$.
Since once $|\varphi(t)\rangle$ is given $\ddot\xi_c(t)$ is linear
in $V$, one can write $V$ as a sum of the pure Coulomb potential
plus a short-ranged potential, evaluate each term separately, and
sum up the results at the end. In what follows we therefore treat
the two cases separately.

\subsection{Short-range potentials}\label{short_range}
In order to keep the expressions from becoming too cluttered, we
introduce the notation
\begin{equation}\label{aion}
\tilde a(t)\equiv \frac{\sqrt[4]{2I_p}}{\sqrt\pi}
a(t)\frac{w(E(t))}{|E(t)|}
\end{equation}
Substituting \eq{Keld111} in \eq{xic} we obtain
\begin{eqnarray}\label{ppint}
\ddot\xi_c(t) &=&\sum _{n,n'} \int dpdp'\tilde
a^*(t_{n'}(p'_x))\tilde a(t_n(p_x)) \times \cr &\times&
e^{\frac{-{p'}_\perp^2}{|E(t_n(p'_x))|\sqrt{2I_p}}}e^{\frac{-p_\perp^2}{|E(t_n(p_x))|\sqrt{2I_p}}}
\times \cr&\times& e^{iS(\mathbf p',t,t_{n'}(p'_x))-iS(\mathbf
p,t,t_n(p_x))}Q(\mathbf p'-\mathbf p)
\end{eqnarray}

We now carry out the integration in \eq{ppint} in the
stationary-phase approximation. To this aim we assume that apart
from the exponentials in \eq{ppint}, the rest of the integrand is
slowly varying in $\mathbf p$. Here it is where we use the
short-range property of $V$, which assures that $Q$ is indeed slowly
varying in $\mathbf p$.

The stationary phase condition in the six-dimensional $\mathbf
p$-$\mathbf p'$ momentum space is identical to the one of the
integral (\ref{xip}): $p_\perp=p'_\perp=0$, whereas $p_x$ and $p'_x$
are obtained by finding all solutions $\bar t_n(t)$ that satisfy
\eq{statp} and using \eq{sad_real} to find the corresponding
$p_x$-s. The result of the integration is
\begin{eqnarray}\label{ppint2}
\ddot\xi_c(t) =(2\pi)^{3/2}\sum _{n,n'} \frac{ \tilde a^*(\bar
t_{n'})\tilde a(\bar t_n)}{( t-\bar t_n)^{3/2}(t-\bar t_{n'})^{3/2}}
\cr \times e^{i\bar S(t,\bar t_{n'})-i\bar S(t,\bar t_n)} Q(\mathbf
A(t_{n'})-\mathbf A(t_{n}))),
\end{eqnarray}
where the dependence of $\bar t_n$ on $t$ has been suppressed.

\eq{ppint2} confirms our claim that $\ddot\xi_c(t)$ is small
compared to $\ddot\xi_1(t)$. $\ddot\xi_1(t)$ [\eq{lewedip3}] has
only a $(2\pi/(t-\bar t_n))^{3/2}$ prefactor (which is
$O(\omega^{3/2})$), whereas for $\ddot\xi_c(t)$ it is $O(\omega^3)$.
Moreover, $\xi_1(t)$ is proportional to $\sqrt{w(E)}$, whereas
$\ddot\xi_c(t)$ is linear in $w(E)$. HHG experiments typically
operate under the condition of small ionization per cycle, which
means $w(E)\ll \omega $. Therefore although $\ddot\xi_c(t)$ is
formally of the same order in $V(\mathbf r)$ as $\ddot\xi_1(t)$,
$\ddot\xi_c(t)$ is more than $O(\omega^2)$ smaller than
$\ddot\xi_1(t)$, and is thus negligible under the conditions
(\ref{allreqs}).

It should be noted that since $\xi_1(t)$ is proportional to
$a^*(t)$, at very high fields, where the ground state is almost
completely ionized in one cycle, $\xi_1(t)$ becomes exponentially
small in the field amplitude \cite{GK}. In this case $\ddot\xi_c(t)$
may become significant. However this regime is of little interest
form the point of view of HHG, since HHG basically disappears under
these operating conditions \cite{GK}.

\subsection{Coulomb potential -- different trajectories}\label{ccCoulDiff}

Now we consider the case $V(\mathbf r)=-1/r$, which leads to
\begin{equation}\label{Qcoul}
Q(\mathbf k)= \ii\sqrt{\frac 2 \pi}\frac{k_x}{k^2}.
\end{equation}
Due to the singularity at $k=0$, the stationary phase approximation
should be now used more carefully when integrating \eq{ppint}. To
this aim we now consider an integral of the form
\begin{equation}\label{monster}
I\equiv\ii\sqrt{\frac 2 \pi}\int d^3p d^3p' e^{\ii\tau\frac{(\mathbf
p-\mathbf p_0)^2}{2}-\ii\tau'\frac{(\mathbf p'-\mathbf
p_0')^2}{2}}\frac{ p_x-p_x'}{|\mathbf p-\mathbf p'|^2}.
\end{equation}
$I$ represents the third line of \eq{ppint}, and the rest of the
integrand is slowly varying and can be added later. \eq{monster} has
two oscillating phase factors, centered at $\mathbf p_0$ and
$\mathbf p_0'$, and the Coulomb potential. $\tau$ and $\tau'$ are
real and positive, and by comparison with \eq{ppint} one can see
that they represent the traveling times of the two trajectories.

Using the convolution and Plancharel's theorems, \eq{monster} is
transformed to
\begin{equation}\label{monsterconv1}
I=\frac{(2\pi)^{3/2}}{\tau^{3/2}\tau'^{3/2}} \tilde I(\Delta p_0,s),
\end{equation}
where
\begin{equation}\label{monsterconv2}
\tilde I(\Delta p_0,s)=\int d^3r \frac{
x\exp(i\frac{sr^2}{2}-i\Delta p_0 x )}{r^3}.
\end{equation}
where $s\equiv 1/\tau'-1/\tau$ and $\Delta p_0 \equiv
p_{0x}-p'_{0x}$. We assumed that the $\mathbf p_0-\mathbf p_0'$ is
parallel to the $x$ axis, since this is the case of interest in
\eq{ppint}.

\begin{figure}[htb]
\centering
\includegraphics[width=9cm]{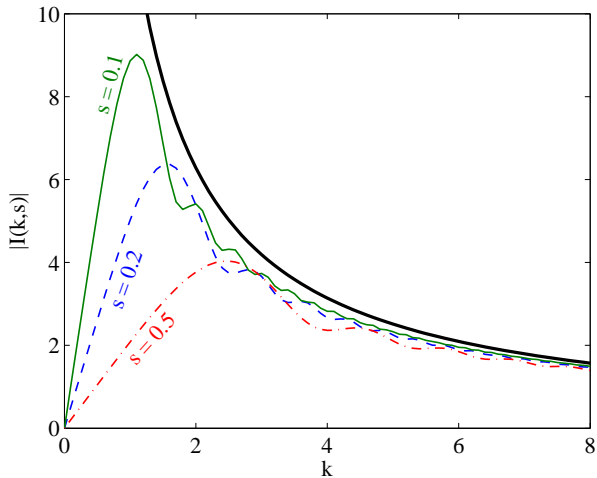}
\caption{$|\tilde I(\Delta p_0,s)|$ [\eq{Isk}] as function of
$\Delta p_0$. The thick line is $4\pi/\Delta p_0$, the limit of
$|\tilde I(\Delta p_0,s)|$ at $s=0$.}\label{erfi}
\end{figure}

The integral in \eq{monsterconv2} can be carried out analytically in
spherical coordinates and expressed in a closed form using the error
function:
\begin{equation}\label{Isk}
\tilde I(\Delta p_0,s)=\frac{4\pi\ii}{\Delta p_0^2}e^{-\frac{\Delta
p_0^2}{2\ii s}}\left(\sqrt{\frac{ \ii\pi s}{2}}{\rm erf}\frac{
\Delta p_0}{\sqrt{2\ii s}}-\Delta p_0\right).
\end{equation}

\eq{Isk} and \eq{monsterconv1} give an exact expression for $I$ in
\eq{monster}. Let us now perform the integration in \eq{monster}
using the stationary-phase approximation instead. The result is
\begin{equation}\label{naive}
I_{\rm SPA}= \frac{\ii2^{7/2}\pi^{5/2}}{\Delta p
\tau^{3/2}\tau'^{3/2}}.
\end{equation}
Since each exponential picks up only an environment of radius
$\tau^{-1/2}$ (or $\tau'^{-1/2}$) around its center $\mathbf p_0$
(or $\mathbf p_0'$), one expects that when $1/\tau,1/\tau' \ll
\Delta p_0^2$, the integration does not reach the singularity and
\eq{naive} holds. \eq{monsterconv1} and \eq{Isk} verify this
expectation. Figure \ref{erfi} visualizes this statement, and shows
that $\tilde I(\Delta p_0,s)$ basically gives a smoothed version of
the $1/\Delta p_0$ singularity near $\Delta p_0=0$.

The discussion in Sec.~\ref{short_range} therefore holds as it is as
long as the condition $1/\tau,1/\tau' \ll \Delta p_0^2$ is met. This
condition is violated when $\Delta p_0$ approaches zero, and this
case will be our concern in what follows. By observing \eq{ppint2}
one can see that the latter always happens when $n=n'$, and can also
accidentally happen if $n\neq n'$. We begin with the second case.

By Taylor-expanding the modulus of \eq{Isk} keeping the two leading
orders in $\Delta p_0$, one can find the \emph{maximal} value of
$|\tilde I(\Delta p_0,s)|$ with respect to $\Delta p_0$, and see
that it is proportional to $s^{-1/2}$. Using this result for an
upper bound on $I$, one obtains
\begin{equation}\label{monsterconv3}
|I|<2^{7/2}\sqrt[4]{\frac{700}{80919}}\frac{\pi^{5/2}
}{\tau\tau'\sqrt{|\tau-\tau'|}}
\end{equation}
Note that we are considering two different trajectories, which means
that $\tau\neq\tau'$. Moreover, it is easy to show that two
different trajectories that return at the same time must have their
birth times separated by more than a quarter of a driving cycle
[$\omega|\tau-\tau'|>\pi/2$]. It follows therefore that
\eq{monsterconv3} is $O(\omega^{5/2})$.

For two different trajectories, $n\neq n'$, the Coulomb singularity
therefore enhances the integral \eq{ppint} by at most a factor of
$\omega^{-1/2}$. This is seen by comparing \eq{monsterconv1} and
\eq{monsterconv3}. In Sec.~\ref{short_range} we have shown that
$\ddot\xi_c(t)$ is smaller than $\ddot\xi_1(t)$ by a factor of
$O(\omega^2)$. Now we arrive at the conclusion that in the long
range case, for two different trajectories, $\ddot\xi_c(t)$ can be
sometimes smaller than $\xi_1(t)$ by factor of $O(\omega^{3/2})$
only. Yet, $\ddot\xi_c(t)$ remains negligible for $\omega\ll 1$.

\subsection{Coulomb potential -- same trajectory} \label{ccCoulSame}
If $n=n'$, which means $\tau=\tau'$ and $\Delta p_0=0$, the integral
(\ref{monster}) diverges. This divergence is an artifact of the
stationary phase approximation and can be avoided if one uses the
fact that the first two lines in \eq{ppint} vanish exponentially at
$p,p'\to\infty$. Doing so is however technically cumbersome, and we
adopt a simpler approach.

We start with having another look at \eq{ppint2}. For a given pair
$n$ and $n'$, the main HHG frequency to be generated is given by the
derivative of the phase in \eq{ppint2}:
\begin{eqnarray}\label{hfreq}
\frac{d}{dt} (\bar S(t,t_{n'})- \bar
S(t,t_n))=\phantom{AAAAAAAAAA}\cr=\frac12
(A(t)-A(t_{n'}))^2-\frac12(A(t)-A(t_{n}))^2
\end{eqnarray}
\eq{hfreq} has a simple intuitive meaning. $\ddot \xi_1(t)$ gives
the beat frequency between the ground state (at frequency $I_p$) and
a continuum electron with a frequency corresponding to the kinetic
energy upon return. In contrast, $\ddot\xi_c(t)$ is obtained from
two different electron trajectories, which return to the parent ion
at the same time with two different kinetic energies. The emitted
radiation is at the beat frequency corresponding to the difference
in the kinetic energies upon return.

Obviously, if $n=n'$, that is, if the two trajectories originated at
the same birth time, they are identical. Therefore the beat
frequency is zero and no high harmonics are generated. In other
words, for $n=n'$ \eq{ppint} varies slowly in time. There is yet a
subtle point to be checked: loosely speaking, if \eq{ppint} is very
large, then even if it varies very slowly in time, its
high-frequency tail may be comparable to $\ddot\xi_1(t)$.

In order to show that this is not the case, we introduce an infrared
cutoff to the Coulomb potential, replacing it by the Yukawa
potential $V(\mathbf r)=-e^{-r/r_0}/r$. After changing the variables
of integration to $\mathbf p_+\equiv \frac12 (\mathbf p+\mathbf p')$
and $\mathbf p_-\equiv \mathbf p-\mathbf p'$, \eq{ppint} has the
form
\begin{equation}\label{monster2}
\ii\sqrt{\frac{2}{\pi }}\int d^3p_+ d^3p_- e^{i\tau\mathbf p_+\cdot
\mathbf p_-}\frac{ p_{-x} g(\mathbf p_+,\mathbf
p_-)}{p_-^2+r_0^{-2}},
\end{equation}
where $g$ represents the slowly-varying part of the integrand.

The integral (\ref{monster2}) is governed by the vicinity of
$\mathbf p_+=\mathbf p_-=0$, and we therefore Taylor-expand $g$
around the origin. The zeroth order term, where $g(\mathbf
p_+,\mathbf p_-)$ is replaced by $g(0,0)$, vanishes since the
integrand is odd in $(\mathbf p_+,\mathbf p_-)$. Since the first two
lines of the integrand in \eq{ppint2} consist of two identical
functions of $\mathbf p$ and $\mathbf p'$, the derivative of $g$
with respect to $\mathbf p_-$ vanishes at the origin. The leading
non-vanishing term in \eq{monster2} comes from the derivative of $g$
with respect to $p_{+x}$, giving:
\begin{eqnarray}\label{monster3}
\ii\sqrt{\frac{2}{\pi }}g_+\int d^3p_+ d^3p_- e^{\ii\tau\mathbf
p_+\cdot \mathbf p_-}\frac{ p_{-x}p_{+x}}{p_-^2+r_0^{-2}}=\cr =
-\frac {g_+}{3\tau^2}\int d^3p_+ \frac{e^{-\tau
p_+/r_0}}{p_+}=-\frac{4\pi g_+ r_0^2}{3\tau^4},
\end{eqnarray}
where $g_+\equiv \partial_{p_{+x}}g(0,0)=-\dot{\tilde a}(\bar
t_n(t))/E(\overline t_n(t)).$

The range of the Yukawa potential, $r_0$, should be set to a value
that is large enough such that the wavefunction is contained within.
Classically, the electron travels to distances of the order of
$E_0/\omega^2$, whereas $\tau$ is of the order of $1/\omega$.
Therefore the modulus of expression in \eq{monster3} is of the order
of $\dot{\tilde a}(\bar t_n(t))E_0$.

We shall not delve into the analysis of $\dot{\tilde a}(\bar
t_n(t))E_0$ any further, being content with the fact that it is much
smaller than one, and is a smooth function of time that varies over
a timescale of $\omega^{-1}$. These two properties assure that
frequencies of $O(1)$ have amplitudes which are exponentially small
in $\omega^{-1}$, and thus can be neglected. Our statement, that the
continuum-continuum term $\ddot\xi_c(t)$ is negligible compared to
the recombination term $\ddot\xi_1(t)$ as far as HHG is concerned,
is now established.

\section{Derivation of \eq{arecold}}\label{diffxhi0}

The zeroth order SFA wavefunction is defined through the equation
\begin{equation}\label{phiveqV}
i|\dot\varphi(t)\rangle=H_V(t)|\varphi(t)\rangle-a(t) E(t)
x|0\rangle.
\end{equation}
We now differentiate $\xi_0$ twice in time, using \eq{phiveqV}. We
obtain:
\begin{eqnarray}\label{xi0dd}
\ddot\xi_0(t)= i\dot Ea^*(t)\langle 0
|x^2|\varphi\rangle-a^*(t)\langle 0 |xH_V(t)^2 |\varphi\rangle,
\end{eqnarray}
where terms including derivatives of $a(t)$ have been dropped, as
they are negligible. Terms that do not contain $|\varphi\rangle$
have been dropped as well, as they cannot contribute to HHG.

If we substitute \eq{Keld111} into \eq{xi0dd}, we obtain an
expression for $\ddot\xi_0(t)$ which is identical to \eq{lewedip3},
with the only difference that the matrix element $\langle 0
|\partial_xV(\mathbf r)|\mathbf p\rangle$ is replaced by
\begin{eqnarray}\label{arecold1}
&& \langle 0 |xH_V^2(t) |\mathbf p\rangle-i\dot E(t)\langle 0 | x^2
|\mathbf
 p\rangle=\cr&=& \left(\frac{p^2}{2}+I_p\right)^2\langle 0 |x |\mathbf
 p\rangle-2E(t)\left(\frac{p^2}{2}+I_p\right)\langle 0 |x^2 |\mathbf
 p\rangle\cr&-&E^2(t)\langle 0 |x^3 |\mathbf p\rangle +E(t)i\langle 0 |xp_x |\mathbf
 p\rangle-i\dot E(t)\langle 0 | x^2 |\mathbf
 p\rangle\phantom{AAA}
\end{eqnarray}

Of the terms on the right hand side of \eq{arecold1}, we argue that
the first one is the most dominant in the regime defined by
\eq{reqs}. In order to show this we note that $\langle 0 | \mathbf
p\rangle $ has a pole at $p^2=-2I_p$. This follows simply from the
asymptotic behavior of $\langle 0|\mathbf r\rangle$ at $r\to\infty$,
which is an exponential falloff at the rate of $\sqrt{2I_p}$.
Therefore for $p^2\gg2I_p$, $\langle 0 | \mathbf p\rangle $
typically falls off like $p^2+2I_p$ to some negative power. This
fact can be used to estimate the modulus of the second term on the
right hand side of \eq{arecold1} as $|E(t)\sqrt{p^2+2I_p}\langle
0|x|\mathbf p\rangle|$ up to a factor of order one. Using $|E|\ll
(2I_p)^{3/2}$ from \eq{reqs} one arrives at the conclusion that the
second term is negligible compared to the first one. Using similar
argumentation and using \eq{reqs} one shows that the first term on
the right hand side of \eq{arecold1} is indeed the dominant one.
\eq{arecold} is thus established.

\end{document}